\documentclass[a4paper,fleqn,10pt]{article}
\usepackage{cite,mcite}
\usepackage{slashed}
\pdfoutput=1
\input{global}

\hypersetup{
  pdfauthor={Shankha Banerjee, Daniel Reichelt, Michael Spannowsky}
  pdftitle={},
  pdfkeywords={SMEFT, Electroweak Corrections}
}
\graphicspath{{./}{Images/}}
\begin{document}
\preprint{ IPPP/24/34 \\ MCNET-24-11 }

\title{Electroweak corrections and EFT operators in $W^+W^-$ production at the LHC}

\author{Shankha Banerjee$^1$, Daniel Reichelt$^2$, Michael Spannowsky$^2$}
\institute{
 $^1$ The Institute of Mathematical Sciences, A CI of Homi Bhabha National Institute, Taramani, 600113 Chennai, India \\
 $^2$ Institute for Particle Physics Phenomenology, Durham University, Durham DH1 3LE, UK
}

\maketitle
\begin{abstract}
We investigate the impact of electroweak corrections and effective field theory operators on $W^+W^-$ production at the Large Hadron Collider (LHC). Utilizing the Standard Model effective field theory (SMEFT) framework, we extend the Standard Model by incorporating higher-dimensional operators to encapsulate potential new physics effects. These operators allow for a model-independent approach to data interpretation, essential for probing beyond the Standard Model physics. We generate pseudodata at the next-to-leading order in quantum chromodynamics and include approximate electroweak corrections. Our analysis focuses on the interplay between these corrections and SMEFT operators at leading order. The inclusion of electroweak corrections is crucial as they can counteract the effects predicted by SMEFT operators, necessitating precise theoretical and experimental handling. By examining $pp \rightarrow W^+W^-$ production, a process sensitive to the electroweak symmetry-breaking mechanism, we demonstrate the importance of these corrections in isolating and interpreting new physics signatures. Our results highlight the significant role of electroweak corrections in enhancing the interpretative power of LHC data and in obtaining reliable constraints on new physics interactions. 
\end{abstract}

\section{Introduction}
\label{Sec::intro}
In the pursuit of deepening our understanding of fundamental interactions and particles, high-energy physics experiments at the Large Hadron Collider (LHC) continue to be a cornerstone of modern physics. As the LHC progresses into its high-luminosity phase (HL-LHC)~\cite{ZurbanoFernandez:2020cco}, the sheer volume of data expected necessitates a sophisticated approach to interpretation that is both model independent and robust. An agnostic methodology based on effective operators offers a promising path to interpreting complex datasets. This approach, crucial for probing beyond the Standard Model (BSM) physics, allows for a comprehensive analysis without relying on specific, potentially unverified extensions of the Standard Model (SM) of particle physics. The effective operator framework is instrumental in disentangling new physics signatures from the SM backgrounds, ensuring that our interpretations remain grounded in the data rather than predisposed theoretical models.

The Standard Model effective field theory (SMEFT) (see Refs.~\cite{Buchmuller:1985jz,Grzadkowski:2010es,Contino:2013kra,Elias-Miro:2013eta,Brivio:2017vri} and the references therein) provides a systematic framework to incorporate potential deviations from the SM predictions. SMEFT extends the SM by including higher-dimensional operators that encapsulate potential new physics effects at energies beyond the direct reach of current experiments. These operators, which respect the gauge invariance of the SM, parameterize deviations in theory observables in terms of coefficients that are fitted using experimental data. This framework ensures consistency with known physics and remains sensitive to a wide array of possible new phenomena.

As the precision of measurements at the HL-LHC enhances, it becomes increasingly crucial to account for electroweak corrections in the analysis~\cite{Bierweiler:2012kw}. The large datasets obtained allow for unprecedented detail in measuring the parameters of the SM and beyond, including those modeled by SMEFT. Electroweak corrections, which include contributions from loops involving $W$-, $Z$-bosons, the Higgs particle, and quarks, play a significant role in these measurements. Crucially, the impact of including electroweak corrections may counteract the effects predicted by SMEFT operators. This interplay necessitates careful consideration of these corrections to accurately isolate and interpret the subtle signatures of new physics embedded within the LHC data. Such precision in the theoretical predictions and their experimental verification is indispensable for advancing our understanding of the fundamental constituents of nature.

In this study, we focus on one process, serving as a standard candle for probing the electroweak sector at the LHC. This process, $pp \to W^+ W^-$, is paramount to the LHC program due to its sensitivity to the electroweak symmetry-breaking mechanism and its pivotal role in the precision tests of the SM. Thus, this process offers a robust platform to observe the interplay between electroweak corrections and the effects of new physics as parameterized by SMEFT operators. The $W^+W^-$ channel has been studied extensively both in the context of the SM as well as beyond~\cite{Hagiwara:1986vm,Falkowski:2014tna,Gehrmann:2014fva,Amar:2014fpa,Butter:2016cvz,Berthier:2016tkq,Biedermann:2016guo,Kallweit:2017khh,Baglio:2017bfe,Chiesa:2018lcs,Grojean:2018dqj,DeBlas:2019qco,Ethier:2021bye,Anisha:2021hgc,Aoude:2023hxv} both for the LHC and the past and future $e^+e^-$ colliders. By examining this process at high energies, we aim to demonstrate how electroweak corrections can counter the modifications introduced by SMEFT operators at leading order (LO), thereby highlighting the critical need for precise theoretical and experimental handling of such corrections. This detailed scrutiny is essential for disentangling potential new physics from SM backgrounds and for enhancing the interpretative power of the LHC's vast datasets.

Our work is structured in the following manner. In Sec.~\ref{Sec::SMEFT} we summarize the importance of the high-energy primaries and discuss the SMEFT operators at play. We discuss how our Monte Carlo events are generated with next-to-leading order (NLO) QCD and electroweak corrections in Sec.~\ref{Sec::EventGeneration}. We validate our SM $W^+W^-$ results against CMS and perform a fast detector simulation in Sec.~\ref{Sec::Analysis}. We perform a careful statistical analysis and show our results in Sec.~\ref{Sec::Results}. Finally, we conclude in Sec.~\ref{Sec::Conclusions}.

\section{The $pp \to W^+ (\ell^+ \nu) W^- (\ell^- \bar{\nu})$ process in dimension-6
  SMEFT}
\label{Sec::SMEFT}

It is possible to probe the effects of the higher-dimensional operators at both low- and high-energy processes. The low-energy probes include several Higgs couplings which are measured from the run-1 LHC data. These processes, with relatively large cross sections, have the main disadvantage of being limited by large systematic uncertainties, which are the norm in hadron colliders. The improvement to such coupling measurements will thus not improve drastically even with the accumulation of more integrated luminosity. On the other hand, high-energy probes can be used to constrain leading-order higher-dimensional operators which can give rise to quadratic growth in the center-of-mass energy ($E$) of some differential distributions in some scattering processes, with respect to SM. If one can measure such energy-growing behaviour carefully, the SMEFT effects may be able to trump the systematic uncertainties~\cite{Englert:2017aqb}. In this section, we follow the notations of Refs.~\cite{Franceschini:2017xkh, Banerjee:2018bio}. In Table~\ref{tab::operators}, we list the operators which contribute to the $pp \to W^+W^-$ channel at high energies. Even though there are many more operators which contribute to this process, however at high energies, the four deformations $Z_{\mu}\bar{u}_{L/R}\gamma^{\mu}u_{L/R}$ and $Z_{\mu}\bar{d}_{L/R}\gamma^{\mu}d_{L/R}$ are isolated in the dimension-6 deformed Lagrangian. In Ref.~\cite{Franceschini:2017xkh}, it was first pointed out that the same four SMEFT directions (termed the ``high-energy primaries") also control the $Zh$~\cite{Banerjee:2018bio}, $Wh$~\cite{Banerjee:2018bio, Banerjee:2019twi} and $WZ$~\cite{Franceschini:2017xkh} production channels. At high energies, these four channels correspond to the pair production of the different components of the Higgs doublet owing to the Goldstone boson equivalence theorem~\cite{Chanowitz:1985hj}. These four processes are thus intertwined through the $SU(2)_L$ symmetry for $E > m_V^2$ ($V=W^{\pm},Z$ boson). At high energies, these four seemingly different processes from the point of view of collider physics, are intricately related. This helps us understand the relation between the pseudo-observables in $W^+W^-$ production (like the charged triple-gauge couplings (cTGCs))~\cite{Franceschini:2017xkh,ElFaham:2024uop} with those in $Vh$ production~\cite{Banerjee:2018bio,Banerjee:2019pks,Banerjee:2019twi}. A combination of the $Zh$ and the $WZ$ productions was first shown in Ref.~\cite{Banerjee:2018bio}. These four channels are also related to the weak-boson fusion channel as shown in Ref.~\cite{Araz:2020zyh}.

\begin{table}[htb]
	\centering
		\begin{tabular}{|c|c|}
			\hline
		SILH basis & Warsaw basis \\	
            \hline
            $\mathcal{O}_W=\frac{ig}{2}(H^{\dagger}\sigma^a \overleftrightarrow{D}^{\mu}H)D^{\nu}W_{\mu\nu}^a$ & $\mathcal{O}_L^{(3)}=(\bar{Q}_L\sigma^a\gamma^{\mu}Q_L)(iH^{\dagger}\sigma^a \overleftrightarrow{D}_{\mu}H)$ \\
            $\mathcal{O}_B=\frac{ig'}{2}(H^{\dagger}\overleftrightarrow{D}^{\mu}H)\partial^{\nu}B_{\mu\nu}^a$ & $\mathcal{O}_L=(\bar{Q}_L\gamma^{\mu}Q_L)(iH^{\dagger}\overleftrightarrow{D}_{\mu}H)$ \\ 
            $\mathcal{O}_{HW}=ig(D^{\mu}H)^{\dagger}\sigma^a(D^{\nu}H)W_{\mu\nu}^a$ & $\mathcal{O}_R^u=(\bar{u}_R\gamma^{\mu}u_R)(iH^{\dagger}\overleftrightarrow{D}_{\mu}H)$ \\
            $\mathcal{O}_{HB}=ig(D^{\mu}H)^{\dagger}(D^{\nu}H)B_{\mu\nu}$ & $\mathcal{O}_R^d=(\bar{d}_R\gamma^{\mu}d_R)(iH^{\dagger}\overleftrightarrow{D}_{\mu}H)$ \\
            $\mathcal{O}_{2W}=-\frac{1}{2}(D^{\mu}W_{\mu\nu}^a)^2$ & \\
            $\mathcal{O}_{2B}=-\frac{1}{2}(\partial^{\mu}B_{\mu\nu})^2$ & \\
            \hline
		\end{tabular}
	    \caption{Table shows dimension-6 operators contributing to the high-energy longitudinal diboson production channels in the SILH and Warsaw bases.}
		\label{tab::operators}     
\end{table}

Coming to the $VV$ and $Vh$ processes ($V=W^{\pm},Z$ boson), the amplitudes scale differently for the different combinations of the longitudinally and transversely polarized gauge bosons when comparing between the SM and BSM contributions. We tabulate the various combinations in Table~\ref{tab::polarisation}~\cite{Franceschini:2017xkh}. In this study, we will focus on the longitudinally polarized $W$-boson pair.

\begin{table}[htb]
	\centering
		\begin{tabular}{|c|c|c|}
			\hline
		& SM & SMEFT \\	
            \hline
            $q_{L/R}\bar{q}_{L/R} \to V_LV_L (V_Lh)$ & $\sim1$ & $\sim E^2/\Lambda^2$ \\ 
            $q_{L/R}\bar{q}_{L/R} \to V_{\pm}V_L (V_{\pm}h)$ & $\sim m_W/E$ & $\sim m_W E/\Lambda^2$ \\
            $q_{L/R}\bar{q}_{L/R} \to V_{\pm}V_{\pm}$ & $\sim m_W^2/E^2$ & $\sim E^2/\Lambda^2$ \\
            $q_{L/R}\bar{q}_{L/R} \to V_{\pm}V_{\mp}$ & $\sim1$ & $\sim1$ \\
            \hline
		\end{tabular}
	    \caption{Scaling factors of the diboson amplitudes for transverse ($\pm$) and longitudinal ($L$) polarizations for the SM and SMEFT scenarios. $\Lambda$ is assumed to be the cutoff scale for new physics.}
		\label{tab::polarisation}     
\end{table}

The BSM Lagrangian in the broken phase can be written as follows~\cite{Franceschini:2017xkh}.

\begin{align}\label{eq:semft-lagrange}
    \Delta \mathcal{L}_{\textrm{BSM}} &= \delta g^Z_{uL} \left[ Z^{\mu} \bar{u}_L \gamma_{\mu} u_L + \frac{\cos \theta_W}{\sqrt{2}} (W^{+\mu} \bar{u}_L \gamma_{\mu} d_L + \textrm{h.c.}) + \ldots \right] + \delta g^Z_{uR} \left[ Z^{\mu} \bar{u}_R \gamma_{\mu} u_R \right] \nonumber\\
    &+ \delta g^Z_{dL} \left[ Z^{\mu} \bar{d}_L \gamma_{\mu} d_L - \frac{\cos \theta_W}{\sqrt{2}} (W^{+\mu} \bar{u}_L \gamma_{\mu} d_L + \textrm{h.c.}) + \ldots \right] + \delta g^Z_{dR} \left[ Z^{\mu} \bar{d}_R \gamma_{\mu} d_R \right] \\
    &+ ig\cos{\theta_W} \delta g_1^Z \left[ Z^{\mu}(W^{+\nu}W_{\mu\nu}^- - \textrm{h.c.})+Z^{\mu\nu}W_{\mu}^+W_{\nu}^- + \ldots \right] \nonumber \\
    &+ ie\delta\kappa_{\gamma}[(A_{\mu\nu}-\tan{\theta_W}Z_{\mu\nu})W^{+\mu}W^{-\nu} + \ldots], \nonumber
\end{align}
with $Z_{\mu\nu}\equiv \hat{Z}_{\mu\nu}-i W^+_{[\mu}W^-_{\nu]}$, $A_{\mu\nu}\equiv\hat{A}_{\mu\nu}$, $W_{\mu\nu}^{\pm} \equiv \hat{W}_{\mu\nu}^{\pm} \pm i W^{\pm}_{[\mu}(A+Z)_{\nu]}$, where $\hat{V}_{\mu\nu}=\partial_{\mu}V_{\nu}-\partial_{\nu}V_{\mu}$, and $\theta_W$ is the Weinberg angle. The `$\ldots$' refers to the Higgs coupling which we do not explicitly consider in this study~\cite{Gupta:2014rxa, Banerjee:2018bio}. In Table 2 of Ref.~\cite{Franceschini:2017xkh} the relations between the high-energy primaries (denoted by $a_q^{(1)}, a_q^{(3)}, a_u$, and $a_d$) are related to the low-energy primaries. For the high-energy primaries, the Warsaw basis~\cite{Grzadkowski:2010es} of dimension-6 SMEFT operators gives four independent couplings as follows.

\begin{equation}
    a_u = 4\frac{c^u_R}{\Lambda^2}, a_d = 4\frac{c^d_R}{\Lambda^2}, a_q^{(1)} = 4\frac{c^{(1)}_L}{\Lambda^2}, \; \textrm{and} \; a_q^{(3)} = 4\frac{c^{(3)}_L}{\Lambda^2},
\end{equation}

where the above Wilson coefficients are the coefficients of the operators in Table~\ref{tab::operators}. The above parameterization only holds for weakly coupled ``nonuniversal" theories which must have a complete set of operators. Example tree-level completions of such ``nonuniversal" theories include models with a heavy $SU(2)_L$ triplet vector boson that are coupled to the left-handed fermionic currents and to the Higgs current~\cite{Franceschini:2017xkh, Banerjee:2018bio}.

It is important to note here that LEP puts strong constraints on the aforementioned operators when they are considered one at a time. However, from Refs. \cite{Franceschini:2017xkh, Banerjee:2019twi}, one can see that the deviations in the $Zf\bar{f}$ ($\delta g^Z_f$) and $Wf\bar{f'}$ ($\delta g^W_f$) are not only functions of the four operators discussed in this work, but also on operators including $\mathcal{O}_{HD}=(H^{\dagger}D_{\mu}H)^*(H^{\dagger}D^{\mu}H)$, and $\mathcal{O}_{HWB}=H^{\dagger}\sigma^a H W_{\mu\nu}^a B^{\mu\nu}$. In a global analysis, $\delta g^Z_f$ and $\delta g^W_f$ also depends on the two TGCs, $\delta g_1^Z$, and $\delta \kappa_{\gamma}$.

\section{Event Generation}
\label{Sec::EventGeneration}

We generate samples for all relevant processes using \Sherpa version 2.2.15
\cite{Sherpa:2019gpd}. See \cite{Campbell:2022qmc, Andersen:2024czj} for 
an overview of other available frameworks. Matrix elements are generated by the internal tools
\Amegic \cite{Krauss:2001iv} and \Comix \cite{Gleisberg:2008fv}, while virtual
QCD 1-loop corrections are evaluated using \Recola \cite{Biedermann:2017yoi}. The
samples are generated at \NLO in QCD using the \MCatNLO method as implemented in
\Sherpa \cite{Hoeche:2011fd, Hoeche:2012yf}. Events are showered using \Sherpa's Catani-Seymour
dipole-based shower (\CSShower) \cite{Schumann:2007mg} and hadronized using the
cluster fragmentation model implemented in the \Ahadic module
\cite{Winter:2003tt, Chahal:2022rid}. The underlying event is simulated in \Sherpa following the
Sj\"ostrand-Zijl model \cite{Sjostrand:1987su}.

We include approximate electroweak (EW) corrections in \Sherpa based on
\cite{Kallweit:2015dum}, which includes infrared subtracted EW 1-loop
corrections as additional weights to the respective Born cross sections. In
those the event weight is calculated based on the expression
\begin{equation}\label{eq:ewcorr_add}
  \mathrm{d}\sigma_\mathrm{NLO,EW_\mathrm{approx}} = \big[B(\Phi)+V_\text{EW}(\Phi)+I_\text{EW}(\Phi)\big]\mathrm{d}\Phi~,
\end{equation}
where  $B$ denotes the Born contribution also entering the uncorrected QCD cross
section, $V_\text{EW}$ is the electroweak virtual corrections at 1-loop
accuracy, and $I_\text{EW}$ is a generalized Catani-Seymour insertion operator
for EW NLO calculations. The latter subtracts all infrared singularities of the
virtual corrections. This of course is a fundamentally arbitrary procedure, but
should provide a good approximation if electroweak Sudakov logarithms are
dominant. We use both \Recola \cite{Actis:2012qn, Actis:2016mpe} and \OpenLoops \cite{Buccioni:2019sur} to provide the
necessary associated virtual contributions. Alternatively to the additive
Eq.~\eqref{eq:ewcorr_add}, approximate corrections can be multiplicative or
exponential to the Born process. We will consider the difference between the
three as a systematic uncertainty of the approximation.

There are three qualitatively different contributions to consider, the signal $pp\to W^+W^-$ production including  SMEFT operators, the irreducible SM $pp\to W^+W^-$ background and backgrounds from other processes that can lead to the same experimental signatures. We will refer to both of the latter two as backgrounds in the next section, but shall describe the SM contribution to $W$-boson pair production, the central process for this study, in more detail below. We hence generate the following samples with the this general setup.

\paragraph{Standard Model $pp\to W^+(l^+\nu)W^-(l^-\nu)$}
We wish to study the production of a $W$-boson pair with opposite charges. This
sample simulating the SM process contribution is hence central to our study. We use as renormalization and factorization scale the sum of the transverse masses of the two $W$ bosons, i.e.
\begin{equation}
  \mu_R^2 = \mu_F^2 = M_{\perp,W^+}^2 + M_{\perp,W^-}^2
\end{equation}
The electroweak input parameters are set in the so-called $\alpha(M_Z)$ scheme,
and we set
\begin{align}
    M_W &= 80.385~\text{GeV},\;\;&M_Z &= 91.1876~\text{GeV}\\
    M_h &= 125~\text{GeV}\;\;&\alpha_\text{QED}^{-1}(M_Z) &= 128.802~.
\end{align}
The $W$ bosons are decayed leptonically in the narrow width approximation smeared by a Breit-Wigner distribution for the mass of the decay system. We assume the measured partial decay width here. The production of $W^+W^-$ is simulated with \MCatNLO accuracy and we include approximate electroweak corrections as described above. Electroweak corrections including the decayed final state have been studied in \cite{Kallweit:2017khh, Brauer:2020kfv}. Electroweak corrections matched to a parton shower have also been studied in the \POWHEG framework \cite{Chiesa:2020ttl}. We validate the approximate electroweak corrections between the inputs from \OpenLoops and \Recola. After the validation, we perform the bulk of our analysis with the \Recola-based option. Note we do not include any gluon-induced contribution in our analysis.

\begin{figure}
  \centering
\includegraphics[width=0.49\textwidth, height=0.2\textheight]{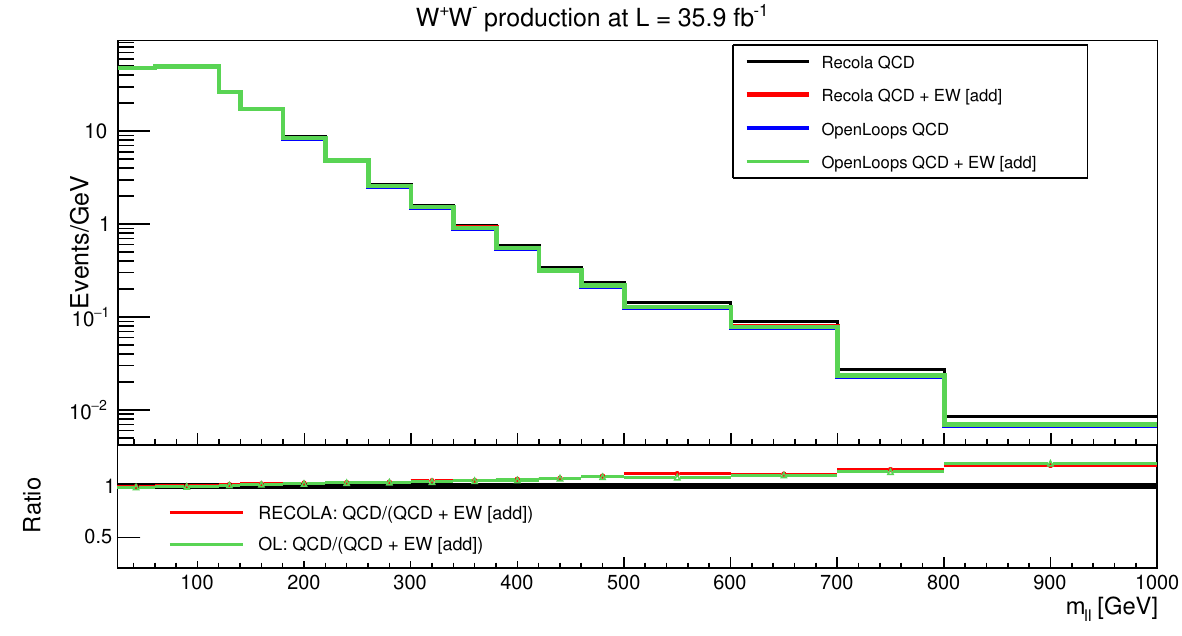}
  \includegraphics[width=0.49\textwidth, height=0.2\textheight]{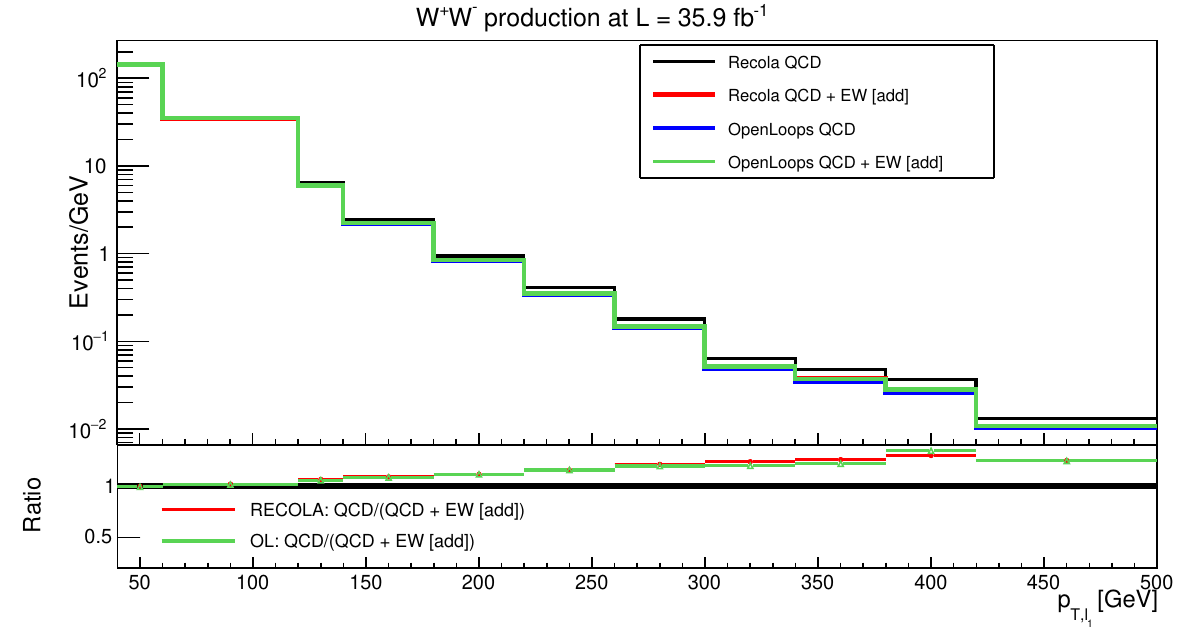}
  \caption{Distribution of the invariant mass of the lepton pair $m_{ll}$ (left) and 
    transverse momentum of the leading lepton $p_{T l_1}$ in $W^+W^-$ production 
    at a center-of-mass energy of $\sqrt{s}=13~\text{TeV}$. Predictions are shown with 
    and without approximate electroweak corrections build from inputs from either 
    \OpenLoops or \Recola}\label{fig:ol_vs_recola}
\end{figure}

To this end, in Fig.~\ref{fig:ol_vs_recola} we illustrate the effect of the
electroweak corrections and compare the two inputs, at the example of the
invariant mass of the final lepton pair and the transverse momentum of the
leading lepton in the event. As expected, electroweak corrections become
important at large scales close to the TeV in the dilepton invariant mass. In both distributions, we observe excellent agreement between \Recola- and \OpenLoops-based approximate corrections, giving us confidence that we can work with just one of the samples for our main analysis.

\begin{figure}
  \centering
  \includegraphics[width=0.49\textwidth, height=0.2\textheight]{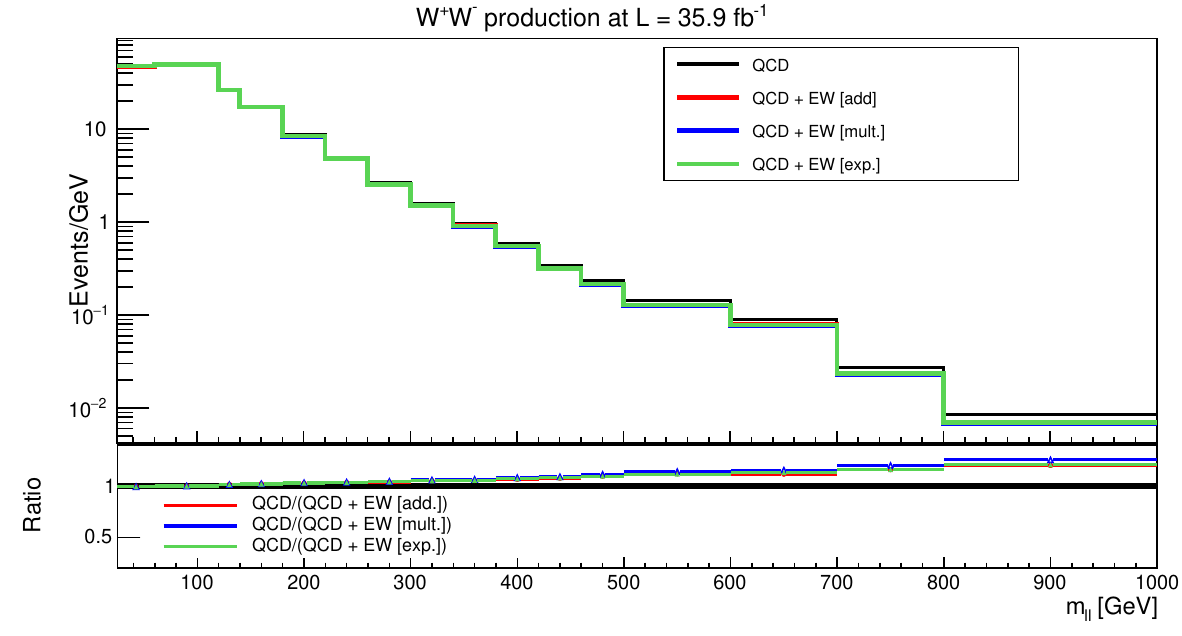}
  \includegraphics[width=0.49\textwidth, height=0.2\textheight]{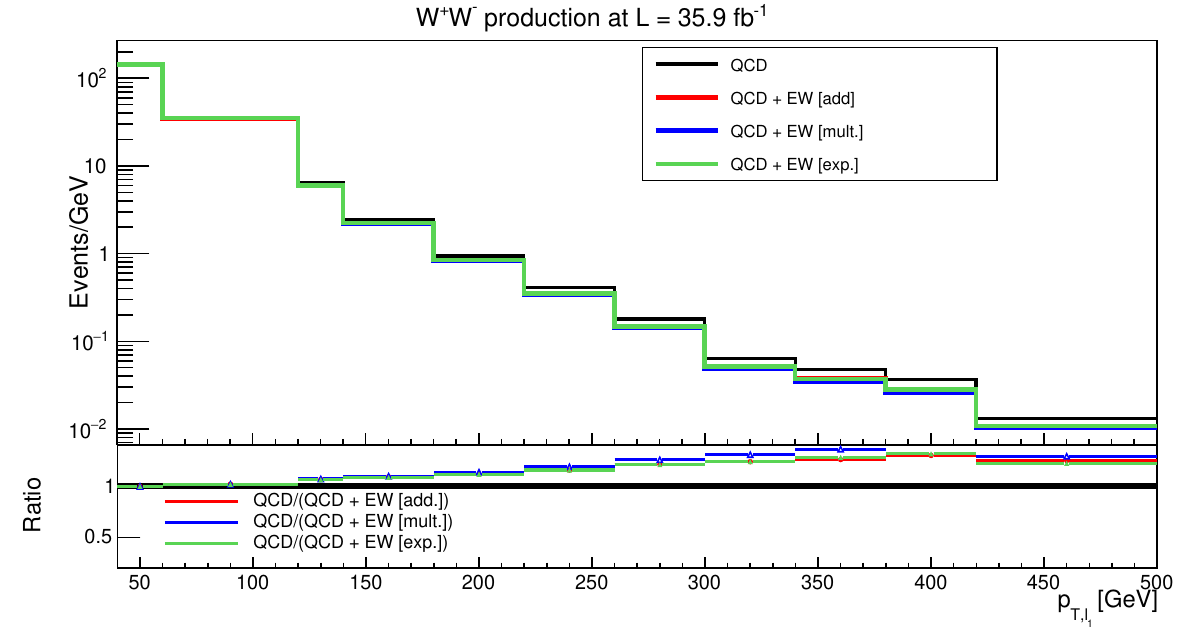}
  \caption{Distribution of the invariant mass of the lepton pair $m_{ll}$ (left) and 
    transverse momentum of the leading lepton $p_{T l_1}$ in $W^+W^-$ production 
    at a center-of-mass energy of $\sqrt{s}=13~\text{TeV}$. Predictions are shown including 
    approximate electroweak corrections in the additive, multiplicative and exponential combination scheme.}\label{fig:qcd_vs_qcdew}
\end{figure}

In order to further investigate the systematics of the approximation, we next
study three different schemes to understand the virtual EW corrections. This is
shown in Fig.~\ref{fig:qcd_vs_qcdew}. We observe small differences between the
schemes, while the qualitative effect remains stable. Quantitatively the effect
of the correction is significantly larger than the uncertainty implied by
comparing the different schemes.

\FloatBarrier

\paragraph{SMEFT+SM interference $pp\to W^+(l^+\nu)W^-(l^-\nu)$}
We use the general UFO \cite{Degrande:2011ua} interface available in \Sherpa
\cite{Hoche:2014kca} to generate samples based on SMEFT in the parametrization
discussed in Sec.~\ref{Sec::SMEFT}. We compute the interference between SM and SMEFT at leading order in both QCD and EW. This is then added, at the level of the final histograms, to the $W^+W^-$ sample described above at NLO QCD and including approximate NLO EW corrections. For the generation of the interference contribution, we assume that SMEFT operators enter in the production of the $W$-boson pair, while for the decays we assume the measured SM branching ratios.
We produce separate samples for one of the couplings in the Lagrangian equation~\eqref{eq:semft-lagrange} in Sec.~\ref{Sec::SMEFT}, $\delta g^Z_{uL}$, $\delta g^Z_{uR}$, $\delta g^Z_{dL}$, $\delta g^Z_{dR}$, set to $1$ while the others are set to $0$. The interference contribution between SM and the SMEFT contribution for the respective part of the Lagrangian is directly proportional to the SMEFT coupling, so the samples can be rescaled to any value of the coupling we need during the analysis, and contributions involving interference of the SM with multiple SMEFT couplings can be obtained by adding the rescaled samples. Note we do not take terms beyond order $1/\Lambda^2$ in the amplitude, so we do not need to produce samples that would involve the product of two or more SMEFT couplings.

\paragraph{Backgrounds}

We further simulate samples for the main backgrounds expected. All backgrounds
are simulated at \MCatNLO accuracy and supplemented with the approximate
electroweak corrections as described above. We mostly follow the default
settings of the \Sherpa Monte Carlo generator, except for when we specify
anything different.  We include a Drell-Yan (DY) sample for $pp\to \ell^+\ell^-$ ($\ell^{\pm} = e^{\pm}, \mu^{\pm}, \tau^{\pm}$) production via a $Z/\gamma^*$ $s$-channel exchange. Vector boson production, including merging of higher jet multiplicities, has been studied
including approximate electroweak corrections and validated against exact
calculations in \cite{Kallweit:2014xda, Kallweit:2015dum} and electroweak
corrections have been studied in \cite{Lindert:2017olm}.  We have separate
samples for the production of vector boson pairs other than $W^+W^-$,
i.e., $WZ$ and $ZZ$ pairs, see \cite{Bothmann:2021led} for a detailed
study of electroweak corrections. $W$ production in association with a lepton
pair is split into the cases where the $W$ decays either hadronically or leptonically. We also include a sample for top pair $pp\to t\bar{t}$ production where electroweak corrections in the above framework have been studied in \cite{Gutschow:2018tuk}. Finally, we consider the production of a top quark together with a $W$ boson, $pp\to tW$. We did not include the $VVV$ processes in our backgrounds as the contributions are at most 0.6\% of the largest background.

\section{Analysis and validation against CMS}
\label{Sec::Analysis}

In order to make our results readily accessible to the readers and to compare against existing experimental analyses, we opt for validating the cut efficiencies for the SM $q\bar{q} \to W^+W^-$ sample. For the validation, we mostly follow Ref.~\cite{CMS:2020mxy}~\footnote{We also heavily rely on a private communication with Guillelmo Gomez Ceballos Retuerto from the CMS collaboration. He is one of the main contacts of the $W^+W^-$ analysis and provided us with crucial efficiency factors for the SM $W^+W^-$ validation.}. For the object selection criteria, we choose isolated leptons and photons with $p_{T,\ell/\gamma} > 10$ GeV, light jets with $p_{T,j} > 30$ GeV, and $b$-tagged jets with $p_{T,b} > 20$ GeV. For the electrons and the photons, we require $|\eta(e/\gamma)| < 2.5$ barring the barrel-endcap region $1.479 < |\eta(e)| < 1.566$ for the electron. For the muons and the $b$-tagged jets, we necessitate $|\eta(\mu/b)| < 2.4$ and for the light jets, $|\eta(j)| < 4.7$. For the tagging and misindentification efficiencies, we choose a ``very loose" working point with $\epsilon_{b\to b} = 81\%$, $\epsilon_{c\to b} = 42\%$, and $\epsilon_{j\to b} = 15\%$~\footnote{These numbers were obtained from a private correspondence with one of the CMS contacts for Ref.~\cite{CMS:2020mxy}.}. The electrons (muons) are considered isolated when the sum $p_T$ of all the particle-flow candidates within a cone radius of $\Delta R = \sqrt{\Delta \eta^2 + \Delta \phi^2} < 0.3 \; (0.4)$ around it and excluding itself is required to be 6\% (15\%) of its $p_T$. For the isolated photon $\Delta R < 0.3$ is chosen and the sum $p_T$ of the particle-flow candidates is required to be less than 10\% of its $p_T$. The remainder of the cuts used in our analysis are tabulated in Table~\ref{tab::cuts}. The variables $m_{\ell\ell}$, $p_{T\ell\ell}$, $p_T^\mathrm{miss}$, and $p_T^\mathrm{miss,proj}$ are, respectively the invariant mass of the two isolated leptons, the combined transverse momentum of the two-lepton system, the missing transverse momentum, and the \textit{projected $p_T^\mathrm{miss}$}. Following Refs.~\cite{CMS:2016lmd} and~\cite{FernandezManteca:2801086}, we identify the isolated lepton closest to $\vec{p}_T^\mathrm{\;miss}$ in azimuthal angle and compute the difference, $\Delta \phi$. When $\Delta \phi < \pi/2$, $p_T^\mathrm{miss,proj} = |\vec{p}_T^\mathrm{\;miss}|\sin{\Delta \phi}|$. For $\Delta \phi \ge \pi/2$, $p_T^\mathrm{miss,proj} = |\vec{p}_T^\mathrm{\;miss}|$.

Following Ref.~\cite{CMS:2016lmd}, we refer to the final states involving either $ee$ or $\mu\mu$ as same-flavor (SF) and the ones with $e\mu$ as different-flavor (DF). Note that most of the cuts are equal for the SF and DF cases except for three cuts; $m_{\ell\ell}$, $|m_{\ell\ell}-m_Z|$, and $\slashed{E}_T$, where $m_Z$ is the central value of the observed mass for the $Z$ boson and $\slashed{E}_T$ is the missing transverse energy. The latter two cuts are extremely effective in reducing the enormous 
DY background.

To validate our setup against CMS, we choose the $q\bar{q} \to W^+W^- \to 2\ell +\slashed{E}_T$ channel. For the validation, we choose the same sets of cuts for $e\mu$, $\mu\mu$ and $ee$ except for the multivariate DYMVA score, which is based on a boosted decision tree\footnote{DYMVA is a boosted decision tree based score which helps in further reducing the large DY backgrounds.}. These cuts are tabulated in Table~\ref{tab::cuts}. While most of our cut efficiencies match within 10\% of CMS, there are a few efficiency factors which are different. The `two tight leptons' selection depends on multiple factors including the identification and isolation efficiencies, the differing lepton isolation in the various jet-multiplicity bins, etc. The softer leptons often ensue from the $\tau$ decays and the $\tau$-impact parameter somewhat reduces the lepton $p_T$. The electron efficiencies are lower than the corresponding muon ones because of the existing backgrounds. Even though at the \textit{reconstruction} level it is very efficient to find electrons and muons, the full electron selection is much tighter because there are stronger backgrounds to start with. In order to reduce the backgrounds associated with the electron events, the electron efficiency becomes much smaller. Thus, for the `two tight leptons' cut, we impose, respetively, scale factors of 0.89, 0.34, and 0.56 for the $\mu\mu$, $ee$, and $e\mu$ final states. The $e\mu$ scaling is obtained by taking a square root of the product of the scale factors for the $\mu\mu$ and $ee$ channels.

As for the jet multiplicities in the 0, 1, and $\ge 2$ jet categories, we get $\sim 65:25:10$ in the respective bins. We corroborate this result with the \texttt{Rivet} routine~\cite{Rivet} provided by the CMS collaboration for the analysis of \cite{CMS:2020mxy}. There is another subtlety that we deal with. There is a sizable fraction of events with no generation-level jets with $p_T < 30$ GeV. However, these are found at the reconstruction level. These include the effects of jet smearing~\cite{CMS:2016lmd}, which we include in our analysis~\footnote{We follow Fig. 38 of Ref.~\cite{CMS:2016lmd}.} and the effect of pileup jets which is not included in our generation. Thus, we rescaled the number of reconstructed jet distribution. We use factors of, respectively, 0.87, 1.38, and 1.78 for 0, 1, and $\ge 2$ jet bins~\footnote{Both the `two tight leptons' scale factors and the jet-bin scale factor were obtained from a private correspondence with Guillelmo Gomez Ceballos Retuerto~\cite{GuillelmoPrivate}.}. In Table~\ref{tab::WW_events}, we show the validation of the $q\bar{q} \to W^+W^-$ channel in the nine final states.

Within our setup, we also validate the DF scenario for the other SM backgrounds. The major backgrounds are listed in Sec.~\ref{Sec::EventGeneration}. In Table~\ref{tab::SM_backgrounds}, we compare the event numbers of the various backgrounds against Ref.~\cite{CMS:2016lmd}.

Here we do not consider the $gg$-initiated one-loop diagrams. For the $W^+W^-$ process, the SM contribution from the $gg$-initiated loop diagrams in the invariant mass range $m_{\ell\ell}$ between 20 and 300 GeV is of the order $\mathcal{O}(10\%)$ of the $q\bar{q}$ cross section~\cite{CMS:2020mxy}. As demonstrated in Ref.~\cite{Banerjee:2018bio}, in a similar analysis for $Zh$, the EFT contributions to the one-loop processes $gg \to Zh$ and $gg \to ZZ$ were subdominant and did not significantly impact the fit results. However, in a future work, we plan to include all remaining backgrounds for completeness, including those stemming from $VVV$ and nonprompt leptons, as well as the EFT effects in the $gg \to W^+W^-$ channel. Additionally, the $gq\bar{q}$ vertices may also be modified by chromomagnetic operators, which are typically strongly constrained by $t\bar{t}$ production. Overall, these effects would be valuable in a global analysis, and we intend to explore them in future work. However, we argue they are not crucial for our goal of demonstrating and estimating the effect of including electroweak corrections into the analysis. The `tops' ($t\bar{t}+tW$) and DY backgrounds show some deviations due to statistical limitations, which are significant due to the large cross sections of these processes. The NLO QCD and NLO electroweak cross section of $t\bar{t}$ is $\sim 700$ pb and that of DY is $\sim 3100$ pb. We do not validate the other SM backgrounds for the SF scenario because we do not employ the multivariate DYMVA analysis as performed in Ref.~\cite{CMS:2016lmd}. In our high-energy analysis, we impose stronger cuts on the $p_T$ of the two leptons and on $m_{\ell\ell}$.

\begin{table}[htb]
	\centering
		\begin{tabular}{|c|c|c|}
			\hline
		Cut & DF & SF \\	
            \hline
            At least two loose leptons & & \\
            Exactly two loose leptons & & \\
            $p_{T_{\ell_1/\ell_2}}$ [GeV] & 20 & 20 \\
            Flavor selection & $e\mu$ & $\mu\mu, $ or $ee$\\
            Two tight leptons & & \\
            Opposite sign leptons & & \\
            $p_{T_{\ell_1}}$ [GeV] & 25 & 25 \\
            $p_{T_{\ell_2}}$ [GeV] & 20 & 20 \\
            $m_{{\ell \ell}}$ [GeV] & 20 & 40 \\
            $|m_{{\ell \ell}}-m_Z|$ [GeV] & $-$ & 15 \\
            $p_{T_{\ell \ell}}$ [GeV] & 30 & 30 \\
            $\slashed{E}_T$ [GeV] & 20 & 55 \\
            $\slashed{E}_T^\mathrm{miss, proj.}$ [GeV] & 20 & 20 \\            
            Number of jets & $\le 1$ & $\le 1$ \\
            Number of $b$-tagged jets & 0 & 0 \\
            \hline
		\end{tabular}
	    \caption{Selection criteria for DF and SF dilepton events following Ref. \protect \cite{CMS:2020mxy}.}
		\label{tab::cuts}     
\end{table}

\begin{table}[htb]
	\centering
		\begin{tabular}{|c|c|c|c|}
		    \hline
            Final state & $0-$jets & $1-$jet & $\ge 2-$jets \\
            \hline
            $e\mu$ [CMS internal] & 6632 & 2953 & 1348 \\
            $e\mu$ [This analysis] & 6816 & 3059 & 1413 \\              
            $e\mu$ \cite{CMS:2020mxy} & $6430 \pm 250$ & $2530 \pm 140$ & NA \\          
            \hline
            $\mu\mu$ [CMS internal] & 5388 & 2332 & 1069 \\
            $\mu\mu$ [This analysis] & 5508 & 2554 & 1175 \\             
            $\mu\mu$ \cite{CMS:2020mxy} & NA & NA & NA \\             
            \hline
            $ee$ [CMS internal] & 2041 & 915 & 434 \\
            $ee$ [This analysis] & 2076 & 922 & 436 \\                
            $ee$ \cite{CMS:2020mxy} & NA & NA & NA \\          
            \hline            
		\end{tabular}
	    \caption{$q\bar{q} \to WW$ events validation with CMS (internal correspondence) and Ref. \protect \cite{CMS:2020mxy}. The integrated luminosity for this validation is $\mathcal{L}=35.9 \; \textrm{fb}^{-1}.$}
		\label{tab::WW_events}     
\end{table}

\begin{table}[htb]
	\centering
		\begin{tabular}{|c|c|c|}
		    \hline
            Background & $0-$jets & $1-$jet \\
            \hline
            Tops ($t\bar{t}+tW$)~\cite{CMS:2016lmd} & $2110 \pm 110$ & $5000 \pm 120$ \\
            Tops ($t\bar{t}+tW$) [This analysis]    & 1678           & 6894 \\
            \hline
            Drell-Yan~\cite{CMS:2016lmd}           & $129 \pm 10$   & $498 \pm 38$ \\
            Drell-Yan [This analysis]              & 14             & 740 \\
            \hline
            $VZ \; (WZ+ZZ)$~\cite{CMS:2016lmd}     & $227 \pm 13$   & $270 \pm 12$ \\
            $VZ \; (WZ+ZZ)$ [This analysis]        & 96             & 183 \\
            \hline
            $W\ell\ell$~\cite{CMS:2016lmd}         & $147 \pm 17$   & $136 \pm 13$ \\
            $W\ell\ell$ [This analysis]            & 105            & 197 \\
            \hline
		\end{tabular}
	    \caption{Comparing number of events for the other SM backgrounds between this analysis and \protect \cite{CMS:2020mxy}. The integrated luminosity for this validation is $\mathcal{L}=35.9 \; \textrm{fb}^{-1}.$}
		\label{tab::SM_backgrounds}     
\end{table}

\section{Results}
\label{Sec::Results}

After validating the $W^+W^-$ SM channel with CMS in Sec.~\ref{Sec::Analysis}, we want to see the effects of including the NLO EW corrections on the four high-energy primaries. We use 1D and 2D $\chi^2$ analyses to show this impact. We take the invariant mass of the two leptons, $m_{\ell\ell}$, as our variable of choice for bounding the four high-energy primaries. Alternatively, we could have used $p_{T_{\ell_1}}$, $p_{T_{\ell_2}}$, or $\slashed{E}_T$. Given the computational cost of filling the very high-energy bins with enough Monte Carlo events, we choose the highest bin as 1 TeV, thus totaling 17 bins of nonuniform bin widths. We refer to these bins as $i$. We also ensure that for the same flavor case, where we employ an additional cut of $|m_{{\ell \ell}}-m_Z| > 15$ GeV, the bin does not split in this range. We have six subcategories, $j$, per bin, which we refer to as $e\mu-0$, $e\mu-1$, $ee-0$, $ee-1$, $\mu\mu-0$, and $\mu\mu-1$, where `0' and `1' refer to the jet multiplicity. For unweighted events, the statistical uncertainty would go as $\sigma^{\textrm{exp. (theo.)}}_{ij,\textrm{stat.}} = \frac{\sqrt{\hat{N}^{\textrm{exp. (theo.)}}_{ij}}}{\hat{N}^{\textrm{exp. (theo.)}}_{ij}}$, where $\hat{N}^{\textrm{exp. (theo.)}}_{ij}$ is the expected (model) number of events in bin $i$ and subcategory $j$. We perform our analysis at $\mathcal{L}=300$ fb$^{-1}$ and at 3 ab$^{-1}$. For the full analysis, we assume a flat systematic uncertainty of 5\%. For unweighted events, the total uncertainty on the SM prediction for bin $i$ and subcategory $j$ is $\sigma_{ij} = \sqrt{(\sigma^{\textrm{exp.}}_{ij,{\textrm{stat.}}})^2+(\sigma^{\textrm{theo.}}_{ij,{\textrm{stat.}}})^2+(\sigma^{\textrm{exp.}}_{ij,{\textrm{syst.}}})^2+(\sigma^{\textrm{theo.}}_{ij,{\textrm{syst.}}})^2}$, where $\sigma^{\textrm{exp. (theo.)}}_{ij,\textrm{syst.}} = 0.05\times \hat{N}^{\textrm{exp. (theo.)}}_{ij}$ is the systematic uncertainty for the expected (model) events for the same bin and corresponding subcategory. For weighted events, which is our case, the relative statistical uncertainty for bin $i$ scales as $\sqrt{\Sigma_k w_k^2}/{\Sigma_k w_k}$, assuming that there is variation of weights within the bin. We perform a $\chi^2$ fit with an assumption that the correlations between the different processes are negligible. Our second assumption is based on the fact that our analysis is a future prediction for luminosities which have not yet been reached. We use the following formula.

\begin{equation}
    \chi^2 = \sum_i \sum_j \frac{[\mathcal{O}^{\textrm{theo.}}_{ij}(p) - \mathcal{O}^{\textrm{exp., SM}}_{ij}]^2}{\sigma_{ij}^2}
\end{equation}

\begin{figure}
  \centering
  \includegraphics[width=0.49\textwidth]{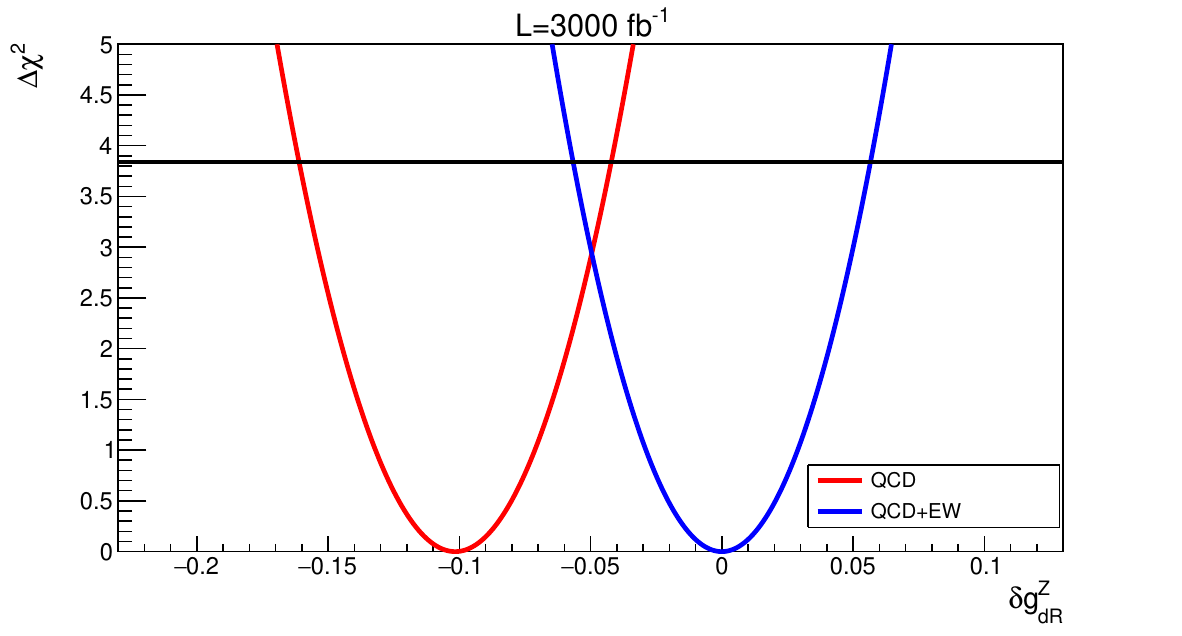}
  \includegraphics[width=0.49\textwidth]{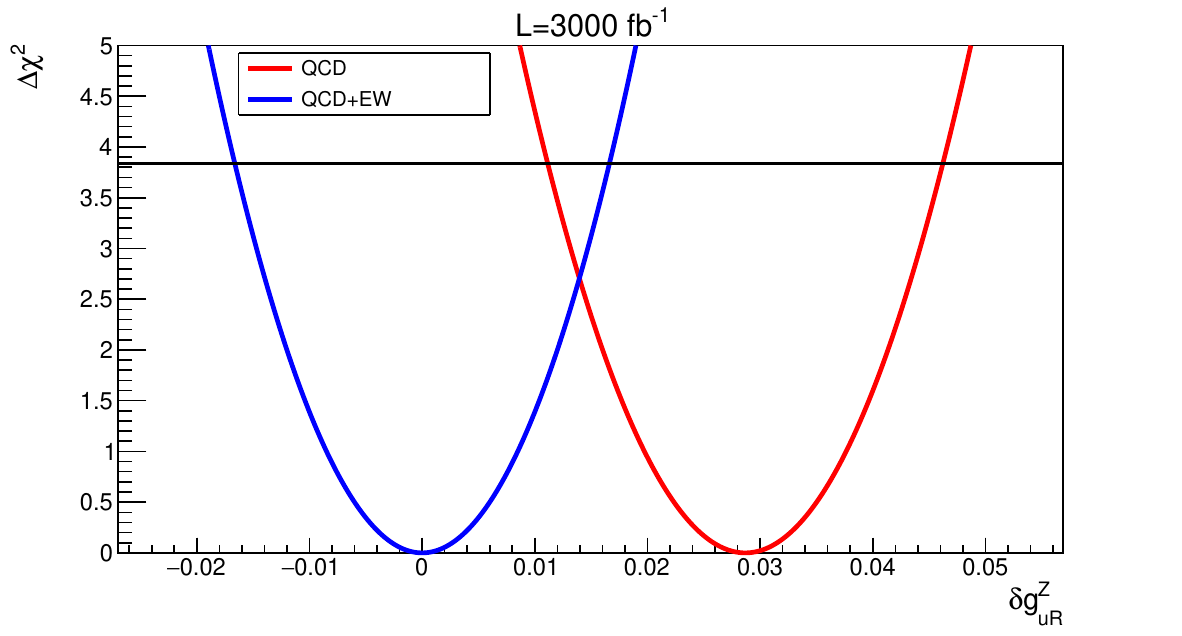}
  \includegraphics[width=0.49\textwidth]{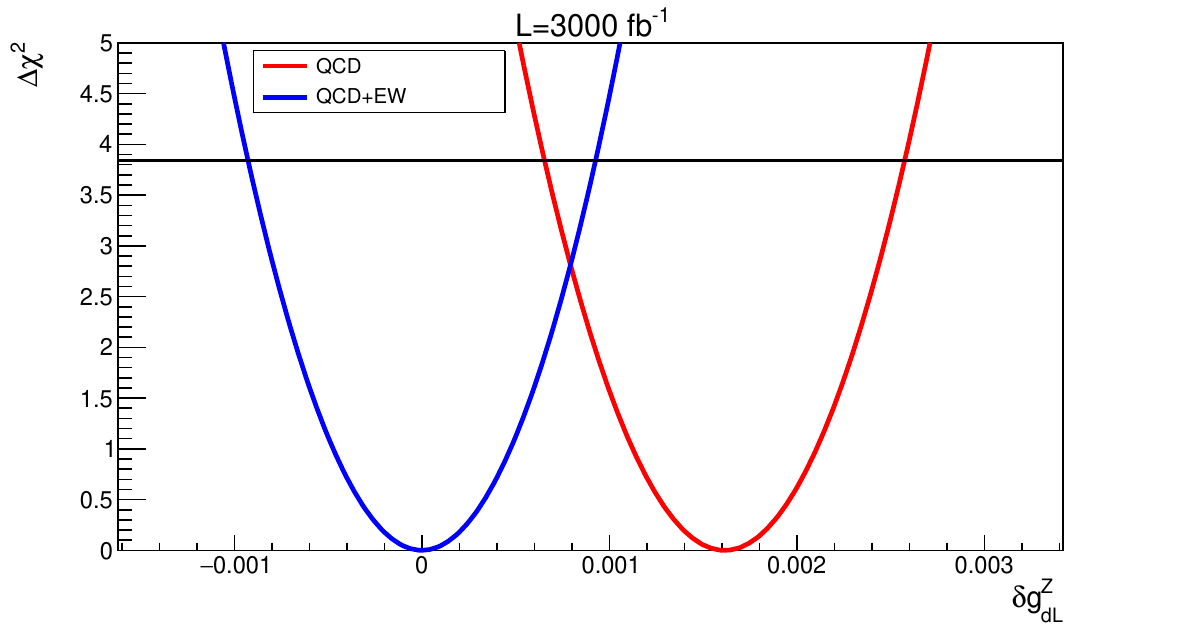}
  \includegraphics[width=0.49\textwidth]{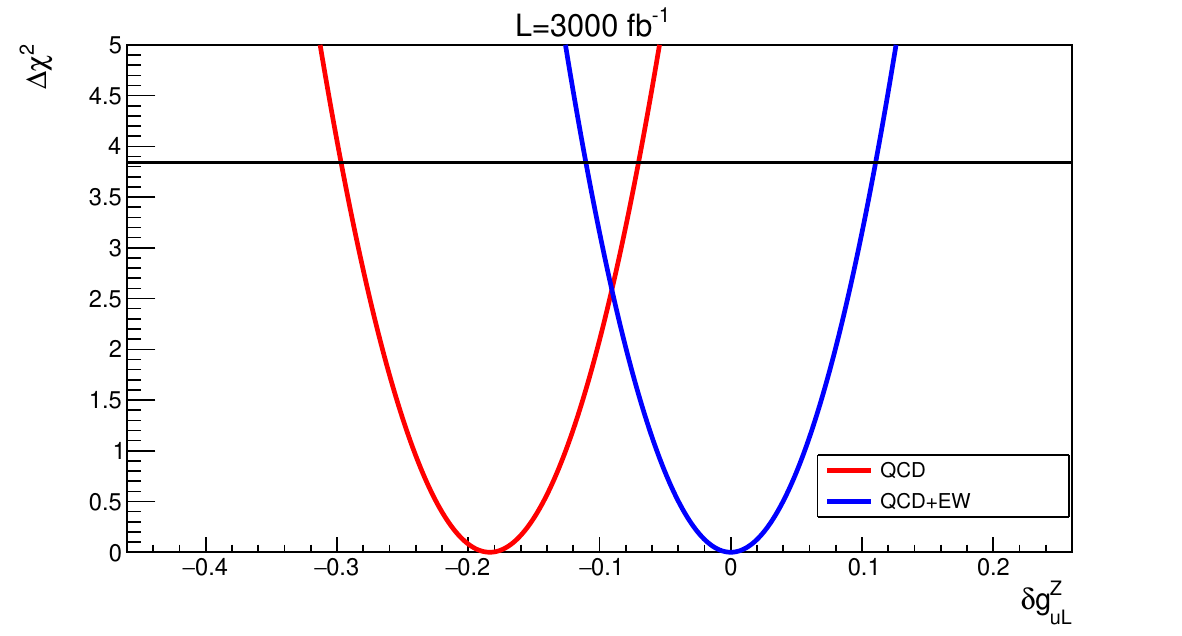}
  \caption{Comparison of the one-dimensional bounds at 95\% C.L. between theory assumptions of SM@(NLO QCD) + SMEFT@LO and SM@(NLO QCD + approximate NLO EW) + SMEFT@LO. The expectation assumption is the same, SM@(NLO QCD + approximate NLO EW) for both theory assumptions. The other parameters have not been marginalized over and we consider $\Delta \chi^2=3.84$ for a one-parameter $\chi^2$ fit to present our allowed regions.}
  \label{fig:1D-bounds}
\end{figure}

\begin{figure}
  \centering
  \includegraphics[width=0.48\textwidth]{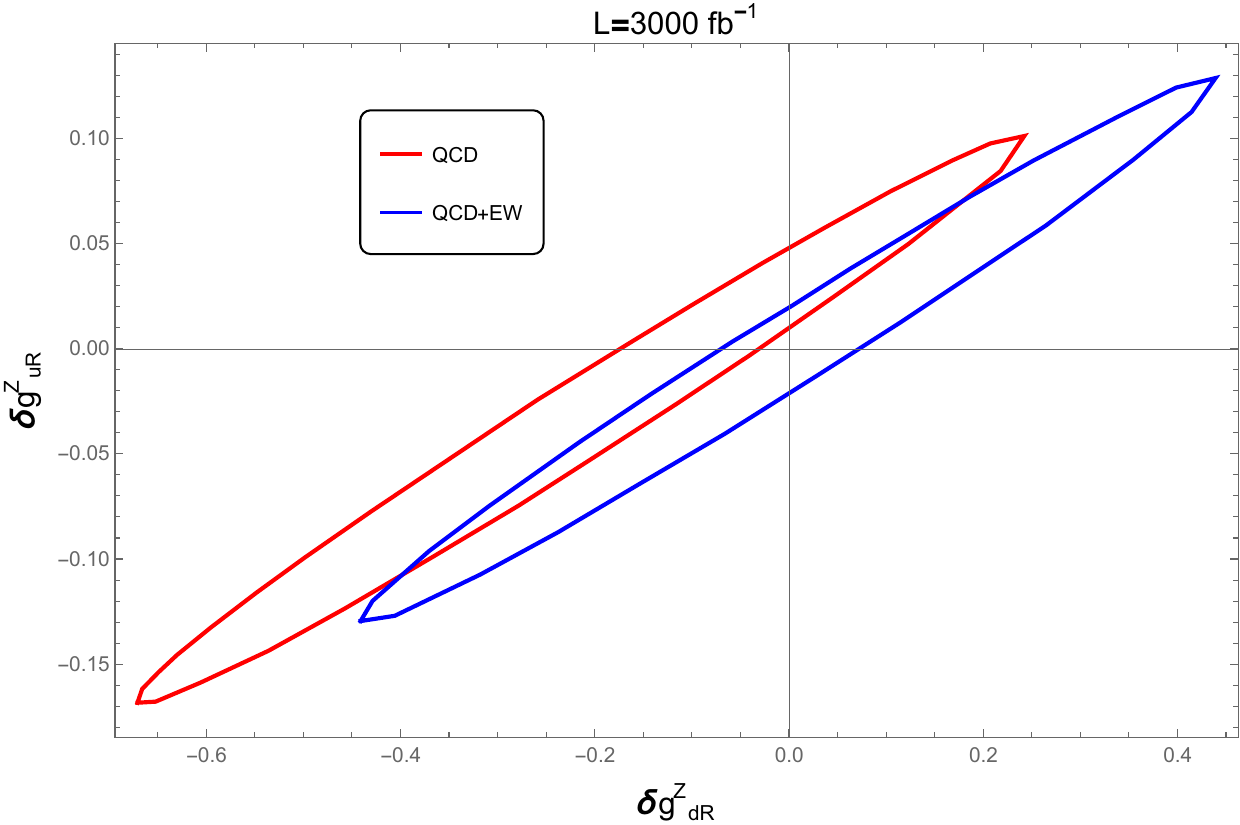}
  \includegraphics[width=0.48\textwidth]{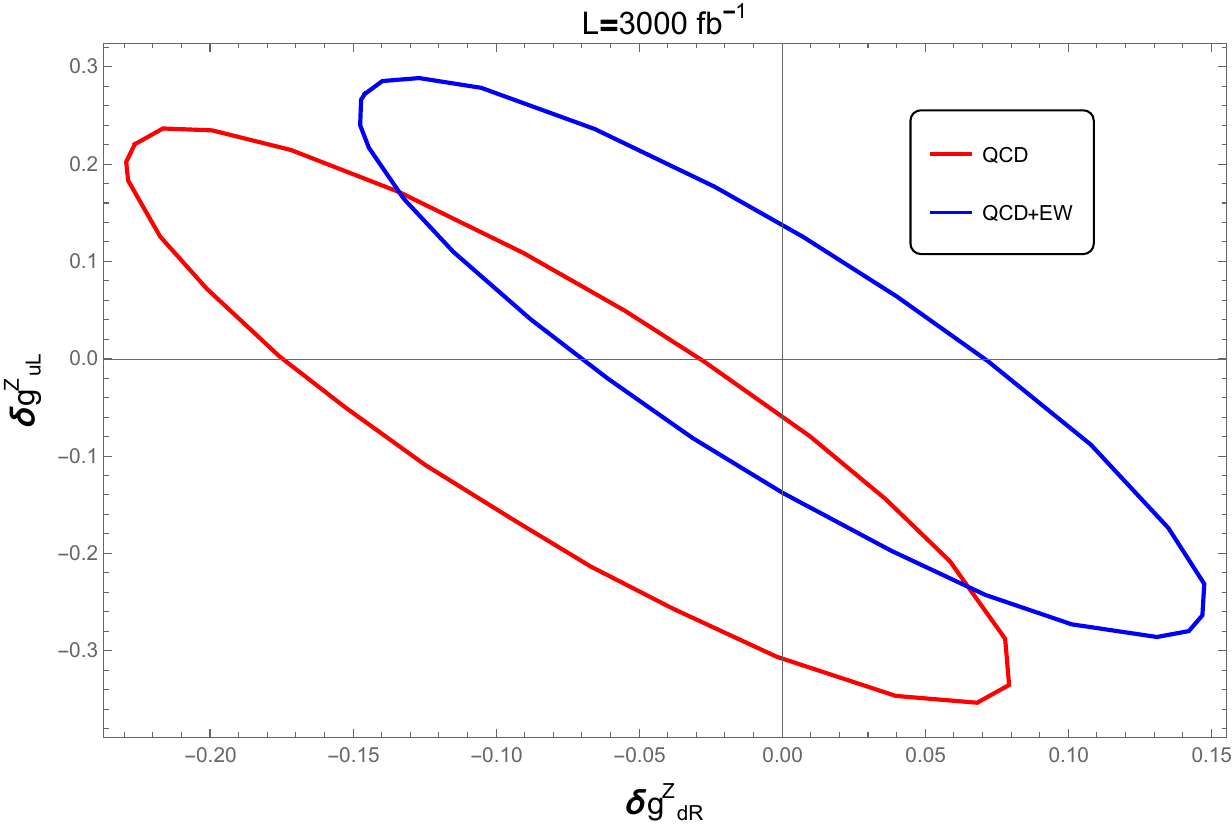} //
  \includegraphics[width=0.48\textwidth]{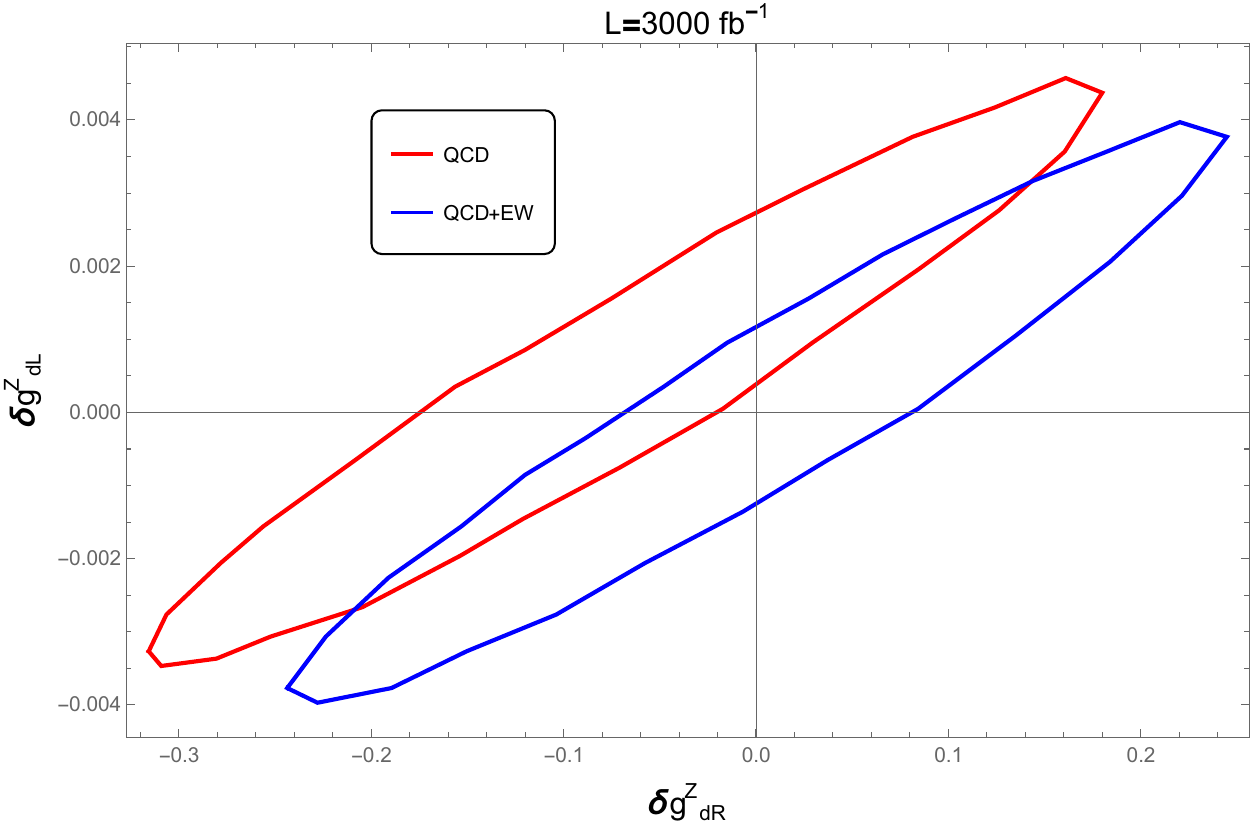}
  \includegraphics[width=0.48\textwidth]{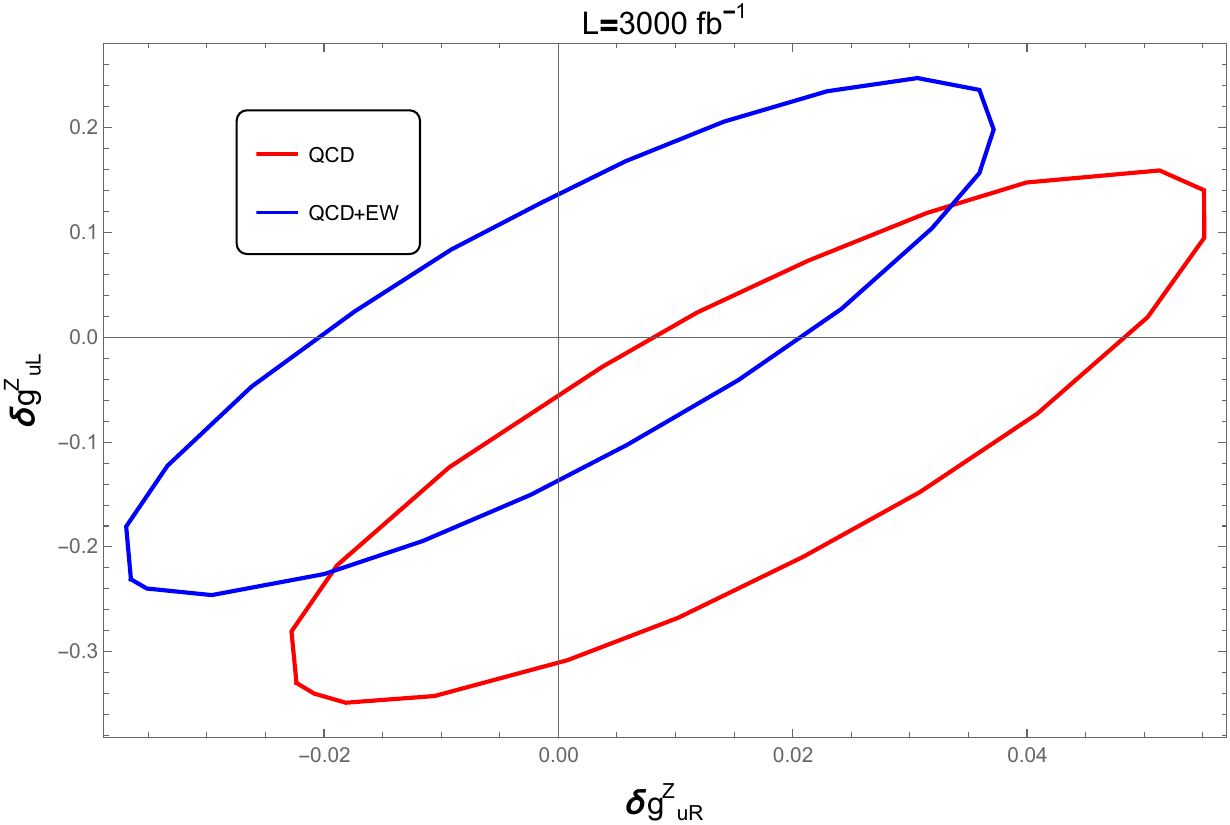} //
  \includegraphics[width=0.48\textwidth]{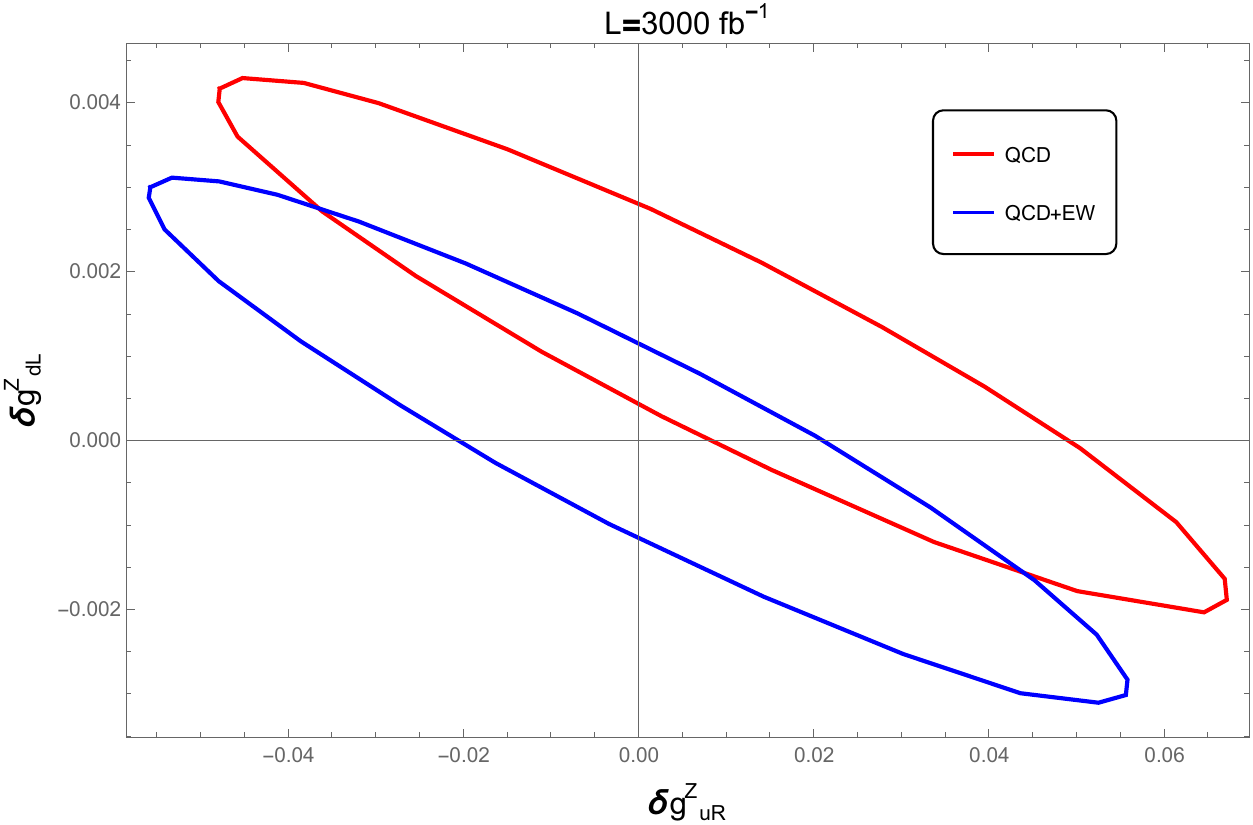}
  \includegraphics[width=0.48\textwidth]{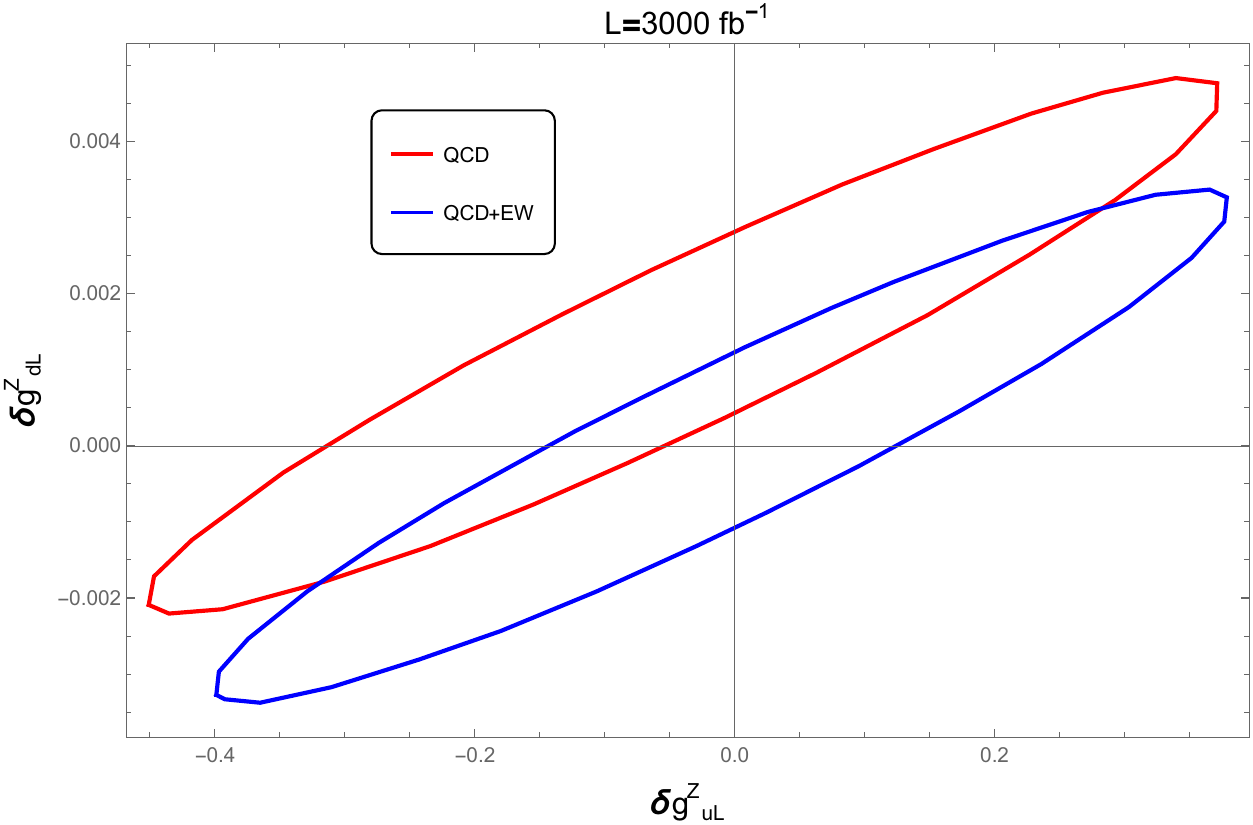} 
  \caption{Comparison of the two-dimensional bounds at 95\% C.L. between theory assumptions of SM@(NLO QCD) + SMEFT@LO and SM@(NLO QCD + approximate NLO EW) + SMEFT@LO. The expectation assumption is the same, SM@(NLO QCD + approximate NLO EW) for both theory assumptions. The other parameters have not been marginalized over and we consider $\Delta \chi^2=5.99$ for a two-parameter $\chi^2$ fit to present our contours.}
  \label{fig:2D-bounds}
\end{figure}

Here $\mathcal{O}^{\textrm{theo.}}_{ij}(p) = \mathcal{O}^{\textrm{SM}}_{ij}+p\times\mathcal{O}^{\textrm{SMEFT}}_{ij}$ is calculated for each bin and each subcategory for a coupling $p=\delta g^Z_{d_R}, \delta g^Z_{u_R}, \delta g^Z_{u_L},$ or $\delta g^Z_{d_L}$. We must note that while considering the statistical and systematic uncertainties ensuing from the `theory' part, we do not consider the contributions coming from the SMEFT pieces, which are the unknowns that are being extracted from the fit. In this analysis, we are only retaining the interference terms under the assumption that the interference pieces dominate over the squared terms. We are thus also not considering the effects of the cross terms between the various operators~\footnote{In this analysis, where the primary goal is to emphasize the importance of including NLO corrections to the SM backgrounds, we retain only the dominant term of the SMEFT signal, namely, the interference term. Due to the Goldstone boson equivalence theorem \cite{Cornwall:1973tb,Vayonakis:1976vz,Chanowitz:1985hj}, the bounds on the four operators at high energies can be translated into constraints on the cTGCs, as discussed in Ref.~\cite{Banerjee:2018bio}. In a similar analysis for $Zh$, the expected constraints at 300 and 3000 fb$^{-1}$ show that the linear term dominates. Furthermore, in certain UV models with top-down matching, the squared term arising from the dimension-6 operator is of the same order as the interference term between the SM and dimension-8 operators. These models are often used to explore the validity region of EFT. A more comprehensive analysis, including the squared contributions, is planned for future work.}. $\mathcal{O}^{\textrm{exp., SM}}_{ij}$ is computed at NLO QCD + approximate NLO EW, as described in Sec.~\ref{Sec::EventGeneration}. This is our pseudodata. In Table~\ref{tab::1D-bounds}, we show the 1D 95\% bounds on the four couplings for $\mathcal{L}=300$ fb$^{-1}$ and 3 ab$^{-1}$. In Fig.~\ref{fig:1D-bounds}, we show the bounds on the four parameters at 95\% C.L. and compare between the following two scenarios: (i) theoretical model is computed at NLO QCD for SM and SMEFT at LO, and (ii) theoretical model is computed at NLO QCD + approximate NLO EW for SM and SMEFT at LO. We see that the bounds are symmetric about zero when the expected theory is taken at NLO QCD + approximate NLO EW for SM and SMEFT at LO. Only the SMEFT interference piece survives and we get no dependence on the sign of the high-energy primaries. On the other hand, when the expected theory only includes SM computed at NLO QCD and SMEFT at LO, the 95\% C.L. bound has a sign dependency as the SM parts of the numerator in the $\chi^2$ function do not cancel out completely. It is seen that $\delta g^Z_{d_L}$ is most strongly constrained. It is important to note that the bound on $\delta g^Z_{d_L}$ is comparable to the one-parameter fit from Ref.~\cite{Banerjee:2018bio}. For the other three high-energy primaries, our bounds are weaker in this analysis. One of the main reasons for that is that we did not venture beyond $m_{\ell\ell} = 1$ TeV~\footnote{In our analysis, we choose $m_{\ell\ell}~(p_{T,\ell\ell})$ in the range $[20, 1000](\sim[30, 500])$ GeV.}. Going to even higher-energy bins may improve these constraints. We checked that the bounds are strengthening when we go to $m_{\ell\ell}$ beyond 1 TeV. However, the main point of this work is to emphasize that the inclusion of the electroweak corrections is imperative. Some of the backgrounds being extremely large in cross section, they are computationally expensive to accurately predict. In Fig.~\ref{fig:2D-bounds}, we show the 2-parameter 95\% C.L. bounds on pairs of the four high-energy primaries.

\begin{table}[htb]
	\centering
		\begin{tabular}{|c|c|c|c|c|}
		    \hline
            Coupling & QCD: $\mathcal{L}=300$ fb$^{-1}$ & QCD+EW: $\mathcal{L}=300$ fb$^{-1}$ & QCD: $\mathcal{L}=3$ ab$^{-1}$ & QCD+EW: $\mathcal{L}=3$ ab$^{-1}$ \\
            \hline
             $\delta g^Z_{d_R}$ & [-0.2744 0.0531] & [-0.1569, 0.1569] & [-0.1611, -0.0421] & [-0.0567, 0.0567] \\
            \hline     
            $\delta g^Z_{u_R}$ & [-0.0180, 0.0818] & [-0.0474, 0.0474] & [0.0111, 0.0463] & [-0.0167, 0.0167] \\
            \hline     
            $\delta g^Z_{d_L}$ & [-0.0008, 0.0039] & [-0.0023, 0.0023] & [0.0006, 0.0026] & [-0.0010, 0.0010] \\
            \hline     
            $\delta g^Z_{u_L}$ & [-0.3910, 0.0927] & [-0.2383, 0.2383] & [-0.2969, -0.0702] & [-0.1104, 0.1104] \\
            \hline                 
		\end{tabular}
	    \caption{The 95\% C.L. bounds from one-dimensional $\chi^2$ fits with the theory prediction to have the SM piece to be computed at QCD or at QCD+EW.}
		\label{tab::1D-bounds}     
\end{table}

Figures~\ref{fig:1D-bounds} and~\ref{fig:2D-bounds} show some crucial features. The model assumption is key in deriving and understanding the constraints on the deformations. Given that the HL-LHC will have excellent statistics and a better handle on systematic uncertainties at the end of its run, the model assumption becomes key in understanding the presence of percent or per-mille level deviations or lack thereof. In this exercise, we find that the constraints on the SMEFT parameters, which can be directly linked to constraints on some UV-complete model's parameters through a top-down matching procedure, depend heavily on our theory assumption. In this case, we can explicitly see that the inclusion of electroweak corrections in the SM part of the theory can drastically alter the allowed regions of new physics. We not only see a shift in the best-fit values, and the allowed ranges, the areas under the two-dimensional contours also change. We see a change between 5 and 9\% for the six scenarios. As SMEFT is used as a tool, the directionality of its deformations would point us towards the nature of new physics. Thus, understanding the higher-order corrections in our theory observables is imperative. Electroweak corrections become as important as QCD corrections in several bosonic processes, especially in regimes where the energies are much higher than the massive gauge bosons. Depending on the sign of the SMEFT interference contribution, the electroweak corrections can change its bounds.

\section{Summary and Outlook}
\label{Sec::Conclusions}

This study investigated the impact of electroweak corrections and EFT operators on $W^+W^-$  production at the LHC. Utilizing the SMEFT framework, we extended the Standard Model to include higher-dimensional operators, providing a model independent approach to probing potential new physics. Our analysis focused on the interplay between electroweak corrections and SMEFT operators at leading order, demonstrating that electroweak corrections can counteract the effects predicted by SMEFT operators. This interplay necessitates precise theoretical and experimental handling to isolate and interpret new physics signatures accurately. We tensioned the outcome of setting constraints on EFT operators when the background is calculated at NLO QCD accuracy against it being calculated at NLO in QCD and ELW accuracy. Our results highlight the significant role of electroweak corrections in enhancing the interpretative power of LHC data and obtaining reliable constraints on new physics interactions. The $pp \to W^+W^-$ process served as a benchmark process for this study. Our validation against CMS data confirmed the accuracy and reliability of our theoretical predictions.

This work pioneers incorporating electroweak corrections into SMEFT analyses, emphasizing their crucial role in high-energy physics. Future studies should continue to include these corrections in similar analyses, such as $Wh$, $Zh$, $WZ$, and weak-boson fusion processes, to ensure precise isolation and interpretation of new physics effects. The following steps should also extend the SMEFT interference piece to include QCD and approximate electroweak effects. Furthermore, additional improvements should involve studying the relevant operators' renormalization group equations~\cite{Englert_2015} and understanding the associated theory systematics. With the high-luminosity Large Hadron Collider providing unprecedented data volumes, the inclusion of electroweak corrections will become even more critical. The precision in theoretical predictions and their experimental verification will be indispensable for advancing our understanding of the fundamental constituents of nature.

\section*{Acknowledgments}
We are grateful to Marek Sch{\"o}nherr for discussions on electroweak corrections in general and in \sherpa specifically. We would like to thank Guillelmo Gomez Ceballos Retuerto from the CMS collaboration for the many detailed exchanges which helped us validate our Monte Carlo samples. S.B. would like to thank Shilpi Jain for her significant help with the ROOT framework, and Rick Gupta for helpful discussions. S.B. would also like to thank the IPPP, Durham University, where this work was initiated, for their  hospitality.

\bibliographystyle{SciPost_bibstyle}
\bibliography{refs}
\end{document}